\documentclass{article}
\usepackage{spconf,amsmath,graphicx}
\usepackage{subfigure}

\title{MoE-AMC: Enhancing Automatic Modulation Classification Performance Using Mixture-of-Experts}
\name{Jiaxin Gao$^{1\;2}$, Qinglong Cao$^{1\;2}$, Yuntian Chen$^{1\;2\;*}$\thanks{* Corresponding author.}}
\address{$^1$ Shanghai Jiao Tong University, $^2$ Ningbo Institute of Digital Twin, Eastern Institute of Technology}
\begin{document}
%\ninept
%
\maketitle
\begin{abstract}
Automatic Modulation Classification (AMC) plays a vital role in time series analysis, such as signal classification and identification within wireless communications. Deep learning-based AMC models have demonstrated significant potential in this domain. However, current AMC models inadequately consider the disparities in handling signals under conditions of low and high Signal-to-Noise Ratio (SNR), resulting in an unevenness in their performance. In this study, we propose MoE-AMC, a novel Mixture-of-Experts (MoE) based model specifically crafted to address AMC in a well-balanced manner across varying SNR conditions. Utilizing the MoE framework, MoE-AMC seamlessly combines the strengths of LSRM (a Transformer-based model) for handling low SNR signals and HSRM (a ResNet-based model) for high SNR signals. This integration empowers MoE-AMC to achieve leading performance in modulation classification, showcasing its efficacy in capturing distinctive signal features under diverse SNR scenarios. We conducted experiments using the RML2018.01a dataset, where MoE-AMC achieved an average classification accuracy of 71.76\% across different SNR levels, surpassing the performance of previous SOTA models by nearly 10\%. This study represents a pioneering application of MoE techniques in the realm of AMC, offering a promising avenue for elevating signal classification accuracy within wireless communication systems.
% CNN-based models, particularly ResNet, excel in high Signal-to-Noise Ratio (SNR) scenarios, whereas Transformer-based models exhibit strong performance in low SNR conditions. Nonetheless, there is a scarcity of models that display robust classification capabilities under both high and low SNR conditions. To address this, we propose MoE-AMC, a Mixture-of-Experts (MoE) based model for AMC. MoE-AMC leverages the advantages of both CNNs and Transformers through the utilization of a gating network that determines the SNR level of input signals. Depending on the determined SNR level, the signals are directed to either the downstream ResNet or the Transformer. 
% We conducted experiments on the RML2018.01a dataset, and our findings demonstrate that MoE-AMC achieves outstanding classification performance under both high and low SNR conditions. MoE-AMC attains an average classification accuracy of 74\% across different SNRs, surpassing the previous SOTA models by over 10\%. This study marks the first application of MoE techniques in AMC, offering a promising solution for improving signal classification accuracy in wireless communication systems.
\end{abstract}
\section{Introduction}
\label{sec:intro}
Automatic Modulation Classification (AMC) is a technology aimed at classifying and identifying signals, which plays a crucial role in the field of wireless communications. A precise and reliable AMC technology enables intelligent signal processing, efficient spectrum resource management, and effective interference control. These capabilities are paramount for enhancing communication system performance, optimizing network resource utilization, and delivering dependable communication services \cite{zhu2015automatic}. However, the task of AMC poses several challenges, including intricate signal patterns, numerous modulation categories, and significant levels of signal noise. 

Deep learning has proven to be an effective technique in AMC due to its ability to automatically learn and extract complex features from raw data. Among the various deep learning models, CNN-based models, especially ResNet \cite{he2016deep}, have been one of the promising models which are successfully applied in AMC \cite{o2016convolutional,o2018over}. The input I/Q dual-channel signals can be reshaped to increase their dimensions, enabling the CNNs to process them in a manner similar to image processing. Leveraging the impressive performance of CNNs in image classification tasks, CNN-based models also demonstrate superior classification performance compared to traditional feature-based methods in AMC. Transformer-based \cite{vaswani2017attention} models have also gained popularity in the field of AMC \cite{hamidi2021mcformer,hu2023feature,chen2022abandon}. Transformers leverage self-attention mechanisms to capture global dependencies and long-range relationships within input signals. Several studies have shown that Transformer-based models exhibit superior performance in classifying low SNR signals \cite{zhang2022modulation,cai2022signal}.  

Nevertheless, the existing models tend to overlook the distinctions between low and high SNR conditions. This oversight leads to an inherent imbalance in tackling the distinct challenges presented by various SNR scenarios. Low and high SNR signals exhibit notable differences in their characteristics. In the case of low SNR signals, the desired signal is significantly attenuated and obscured by noise, making it challenging to extract and decipher the underlying information. The amplitude of the signal is often close to or even below the noise floor, resulting in a weakened SNR. As a consequence, low SNR signals tend to suffer from decreased clarity, reduced distinguishability of features, and an increased susceptibility to interference and distortion. On the other hand, high SNR signals encounter relatively minimal noise contamination, allowing the desired signal to be more discernible and prominent. The signal strength is significantly higher compared to the noise level, enabling clearer and more distinguishable patterns and features. This facilitates effective signal extraction, characterization, and classification. It is imperative to acknowledge the inherent dissimilarities in handling low and high SNR signals and develop a more balanced model that effectively captures the characteristics of both types of signals. By bridging this gap, it is possible to improve the overall performance of classification models across diverse SNR conditions.

The most direct strategies to address the training imbalance arising from different SNRs is to develop three distinct models: one dedicated to SNR level evaluation, another specialized in the classification of high SNR signals, and a third tailored for low SNR signal classification. Our experimental endeavors substantiate the viability of this method. Nevertheless, several issues arise from this method: Firstly, it necessitates the training of three separate models, which can engender computational expenses; secondly, additional labels must be assigned to the input signals to facilitate the training of the model responsible for determining the SNR of signals, although acquiring noise ratio information for signals may not be universally available; finally, a mismatch exists between the training and inference structures, potentially resulting in supplementary costs during deployment.

To tackle these problems, we propose MoE-AMC (Mixture-of-Experts for Automatic Modulation Classification). Mixture-of-Experts (MoE) is a machine learning framework that combines the expertise of multiple specialized submodels, known as experts, to improve the global model's performance and flexibility \cite{chen2022towards, masoudnia2014mixture}. In MoE-AMC, a Multi-Layer Perceptron (MLP)-based gating network is employed to determine the SNR level. Downstream of the gating network, a combination of HSRM (High SNR Recognition Model, a ResNet-based model) and LSRM (Low SNR Recognition Model, a Transformer-based model) is integrated. HSRM demonstrates efficiency in handling and classifying high SNR signals, thanks to its localized perception capabilities, adeptness in capturing pivotal features through pooling operations, and its ability to abstract multi-level features. On the other hand, LSRM is particularly adept at handling low SNR signals. With the self-attention mechanism, LSRM assigns higher importance to relevant features while effectively filtering out noise. It focuses on significant features and suppresses irrelevant noise-induced variations, generating robust feature representations. 
% Unlike CNN-based models that rely on fixed-size receptive fields, Transformer-based models process the entire signal without assuming a predefined temporal structure, allowing them to adapt to varying signal characteristics and capture modulations even in the presence of noise.
When the gating network determines that the input signal belongs to the high SNR category, it is passed to the HSRM; otherwise, it is directed to the LSRM. 

To the best of our knowledge, this study represents the first application of MoE techniques in AMC, enabling deep learning models to achieve excellent modulation classification performance under both high and low SNR conditions. This method partially alleviates the need for denoising preprocessing in identifying high-noise modulation signals. MoE-AMC achieves an average classification accuracy of 71.76\% on the RML2018.01a dataset across different SNRs, surpassing the performance of previous models which achieved less than 63\% accuracy. Importantly, MoE-AMC is not limited to radio signals and can be applied to arbitrary time series data, and the MoE framework employed in this study is not only applicable to time series classification problems with various noise conditions, but can also be extended to time series forecasting, anomaly detection, and other time series-related tasks involving diverse noise conditions. 
% Furthermore, this MoE framework has the potential to handle multiple types of time series data simultaneously, making it a versatile and efficient tool for various applications.

\section{Method}
In this section, we first present the two base models employed in our proposed MoE-AMC: HSRM and LSRM. Subsequently, we outline the overall architecture of MoE-AMC.
\label{sec:method}
\subsection{HSRM for High SNR AMC}
HSRM (High SNR Recognition Model) is designed based on ResNet tailored for high SNR signal classification. ResNet has been widely adopted in various computer vision tasks \cite{he2016deep,dai2016r} and has demonstrated remarkable effectiveness for AMC in scenarios characterized by high SNR, where the input signals exhibit minimal or negligible noise interference. The fundamental concept behind ResNet is its residual structure, which incorporates skip connections to enable smooth information flow throughout the model. This concept can be expressed as: 
\begin{equation}
 f(x) = x + g(x)
 \end{equation}
 where $x$ represents the input, $f(x)$ represents the output of the residual structure, and $g(x)$ denotes the learned mapping or transformation within the residual structure.
% This architectural design effectively mitigates issues related to vanishing gradients and promotes gradient propagation, enabling ResNet architectures to effectively capture and learn intricate features, ultimately leading to superior classification performance.
\begin{figure}[ht]
\centering
\includegraphics[width=0.8\linewidth]{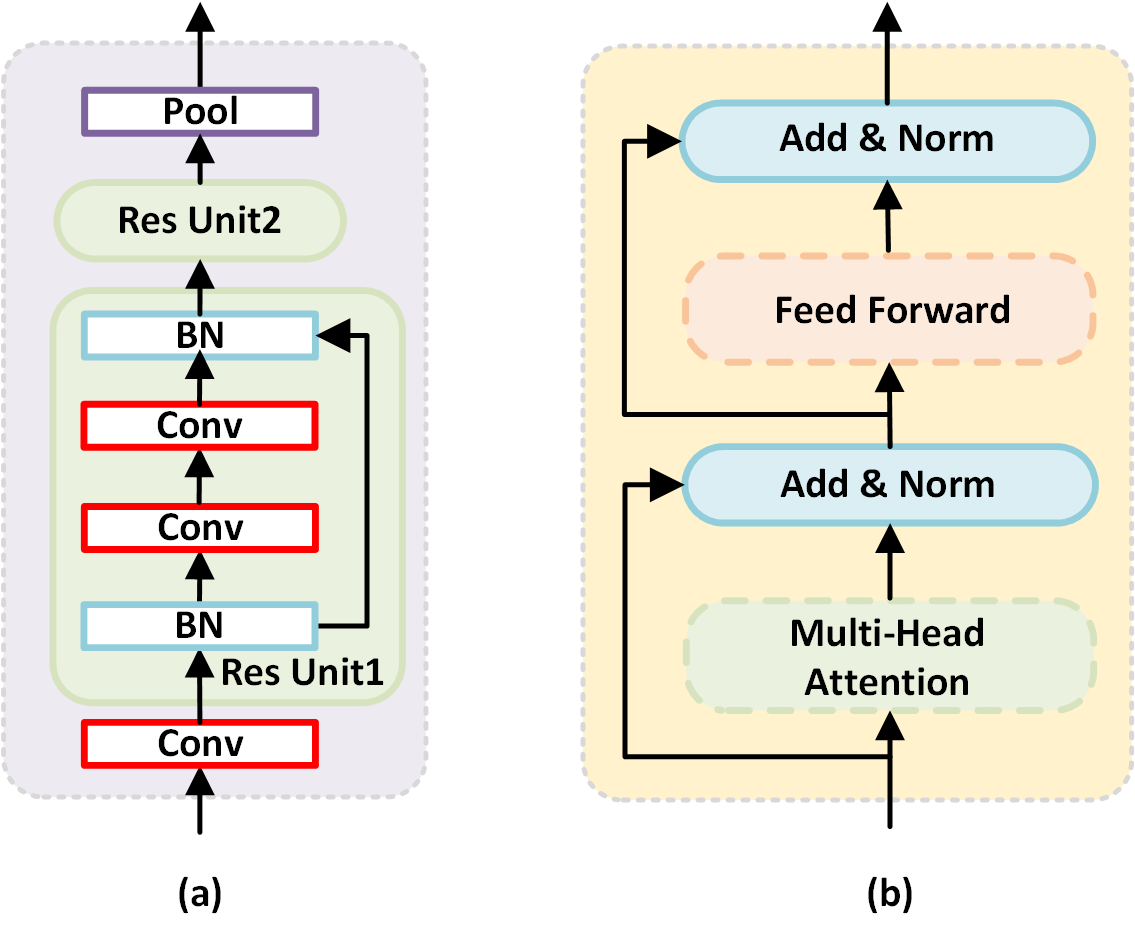} % Reduce the figure size so that it is slightly narrower than the column.
\vspace{-0.5cm}
\caption{(a): The internal structure of the residual stack in HSRM for high SNR AMC. (b): The internal structure of the encoder block in LSRM for low SNR AMC.}
\vspace{-0.3cm}
\label{architecture}
\end{figure}

\begin{figure*}[!t]
\centering
\includegraphics[width=0.8\linewidth]{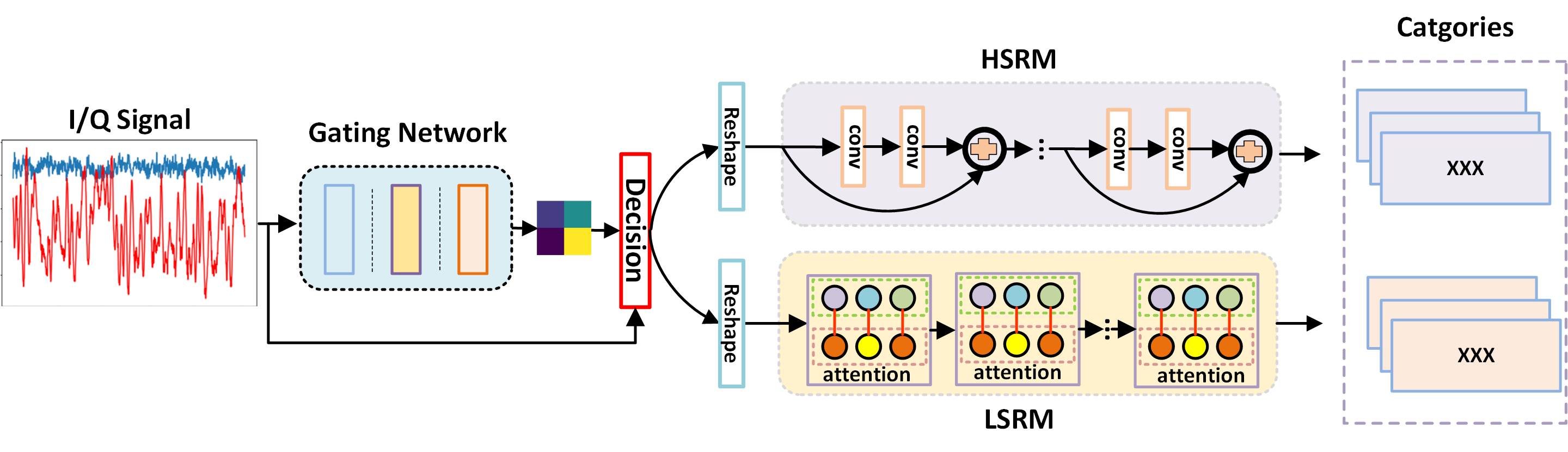} % Reduce the figure size so that it is slightly narrower than the column.
\vspace{-0.5cm}
\caption{The architecture of MoE-AMC. A signal first undergoes SNR evaluation in a gating network, followed by automatic routing to either HSRM (High SNR Recognition Model) or LSRM (Low SNR Recognition Model) based on SNR levels.}
\vspace{-0.5cm}
%The grey boxes in the figure represent the sublayers of the model, and the numbers within the boxes indicate the output shapes of these sublayers. Specifically, "Full" denotes a linear layer, "ReStk" represents the residual stack, and "TransBlock" corresponds to the Transformer encoder block. For the sake of simplicity, certain activation functions or layers such as "Dropout" have been omitted from the figure.} 
\label{fig:MoE-AMC}
\end{figure*}

In the ResNet-based model HSRM, the input signal undergoes an initial reshaping process and is subsequently passed through four residual stacks, each comprising two residual units and a pooling layer, as shown in Fig. \ref{architecture}~(a). The residual units enable the model to learn residual mappings that capture nuanced details and variations present in the input signal, while the pooling layers reduce spatial dimensions while retaining salient features. Following feature extraction, the obtained features are flattened and propagated through three linear layers, with the final linear layer being accompanied by a softmax activation function, yielding classification probabilities for different modulation categories. 

\vspace{-0.4cm}
\subsection{LSRM for Low SNR AMC}
LSRM (Low SNR Recognition Model) is designed based on Transformer for the purpose of classifying low SNR signals. Transformer is originally proposed for NLP tasks \cite{devlin2018bert,brown2020language} and has gained significant attention in the field of AMC, particularly in scenarios characterized by low SNR. The Transformer architecture, based on the self-attention mechanism, excels at capturing long-range dependencies and modeling sequential data. The self-attention mechanism is defined as:
\begin{equation}
\label{equ:attn}
    \operatorname{Attention}(\mathbf{Q}, \mathbf{K}, \mathbf{V} )=\operatorname{softmax}\left(\frac{\mathbf{Q} \mathbf{K}^\top} {\sqrt{d_k}}\right) \mathbf{V}  
\end{equation}
where $\mathbf{Q}$ is queries, $\mathbf{K}$ is keys, and $\mathbf{V}$ is values. $\mathbf{Q},\mathbf{K}$ and $\mathbf{V}$ are generally obtained by applying some transformations to the original input, and ${d_k}$ is the dimension of keys. By effectively encoding temporal information and learning global context, Transformers have demonstrated promising performance in handling noise-corrupted signals. 

In the Transformer-based model LSRM, the input signal also undergoes an initial reshaping process and is subsequently passed through an encoder block to extract features and reduce noise. This encoder block consists of a multi-head attention module and a feed-forward neural network (FFN), interconnected through residual connections and layer normalization, enabling effective information propagation and regularization, as depicted in Fig. \ref{architecture}~(b). Following the feature extraction stage, the model employs a pooling layer for feature compression. The compressed features are then propagated through three linear layers, with a softmax activation function accompanying the final linear layer, yielding the classification results. Unlike HSRM that relies on fixed-size receptive fields, LSRM models the entire signal without assuming a predefined temporal structure, allowing them to adapt to varying signal characteristics and capture modulations even in the presence of noise.

\subsection{MoE-AMC Architecture}

\begin{figure*}[!t]
\centering
\includegraphics[width=0.8\linewidth]{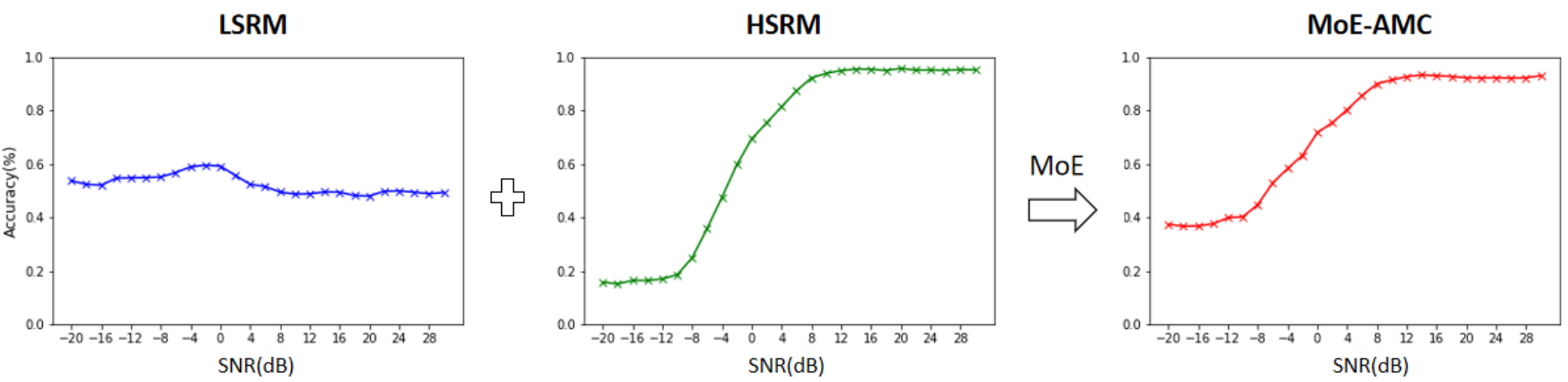} % Reduce the figure size so that it is slightly narrower than the column.
\vspace{-0.4cm}
\caption{The classification accuracies of MoE-AMC and its two base models LSRM and HSRM, at various SNRs.}
\vspace{-0.5cm}
\label{MoE-AMC}
\end{figure*}

MoE-AMC integrates the HSRM and LSRM models within a Mixture of Experts (MoE) framework. When a signal is input into the model, it first undergoes evaluation by a gating network to determine the level of SNR. We employ a Multi-Layer Perceptron (MLP) model consisting of three linear layers (with the final layer accompanied by a sigmoid activation function) to perform this SNR evaluation. HSRM and LSRM, serving as expert models, are connected after the gating network. Once the SNR evaluation is completed, signals with high SNR automatically flow into the HSRM branch, while signals with low SNR automatically flow into the LSRM branch. The overall architecture of MoE-AMC is shown in Fig. \ref{fig:MoE-AMC}.

Maintaining differentiability during the training process is a challenging aspect in the model design. Utilizing a simple if-else statement for signal classification would result in a non-differentiable training process. Our model addresses this issue by designing the gating network MLP to output the probability $y_{high}$ that indicates the likelihood of a signal belonging to the high SNR category. The signal is then separately input into the HSRM to obtain $y_{hsnr}$ and into the LSRM to obtain $y_{lsnr}$. The final output is computed as:
\begin{equation}
 y_{final} = y_{high} * y_{hsnr} + (1 - y_{high}) * y_{lsnr}
 \end{equation}
This solution allows for smooth training of the model while ensuring differentiability. The experimental results indicate that there is no requirement for additional explicit high/low SNR labels or the utilization of extra loss functions. Simply providing the modulation category labels and employing the conventional cross-entropy loss function at the final stage are adequate for MoE-AMC to effectively accomplish the tasks of SNR evaluation and modulation category classification.

\section{Experiment}
\label{sec:exp}
\subsection{Dataset and Experiment Setting}
The experiment utilize the RML2018.01a dataset \cite{o2018over} from GNU Radio ML, which consists of a total of 2,555,904 input samples, with each sample associated with a specific modulation scheme at a given SNR. This dataset encompasses 24 modulation categories, and each category contains an equal number of samples. These samples are generated across 26 different SNR levels ranging from -20dB to 30dB, with an incremental step size of 2dB. Each individual sample has a dimensionality of 2,048, comprising 1,024 in-phase and 1,024 quadrature components.

Several previous SOTA models are adopted for comparison, including MCformer \cite{hamidi2021mcformer}, LSTM \cite{rajendran2018deep}, PET-CGDNN \cite{zhang2021efficient}, MCLDNN \cite{xu2020spatiotemporal}, FEA-T \cite{chen2022abandon}, and two base models of MoE-AMC: LSRM and HSRM.

In the experiments, the dataset is divided randomly into three sets: 70\% of the samples are assigned to the training set, 10\% to the validation set, and the remaining 20\% to the testing set. The training batch size is 1024,  and the training epoch is set to 500, with an early stop function of patience 30. For the training process, the widely used cross-entropy loss function for classification problems is employed. The Adam optimizer \cite{DBLP:journals/corr/KingmaB14} is utilized to optimize the loss function. The evaluation metric for the models is the accuracy obtained across all SNR levels within the testing set. All experiments are conducted on a single NVIDIA Tesla A100 GPU with 80GB of memory.
\vspace{-0.3cm}
\subsection{Result}
The results of AMC performed by different models are presented in Table \ref{result}. The results of MCformer, LSTM, PET-CGDNN, MCLDNN, and FEA-T are adapted from the study \cite{chen2022abandon}. On the other hand, the results of HSRM, LSRM, and MoE-AMC are obtained through our own implementations of these models. It is worth mentioning that these AMC models are derived by modifying or combining conventional deep learning models. Therefore, we have also listed the foundational models used in each AMC model in the table. 

The average accuracy of MoE-AMC across all SNRs reaches 71.8\%, which represents an improvement of nearly 10\% compared to the previous SOTA models. Notably, this performance enhancement primarily stems from the improved classification effect under low SNR conditions, as the previous models have already achieved satisfactory performance under high SNRs. Additionally, our implemented HSRM demonstrates superior performance compared to the previous ResNet-based model proposed in the study \cite{o2018over}. This improvement can be attributed to the incorporation of operations such as increasing the number of convolutional kernels and adding batch normalization layers.

\begin{table}[!htbp]
 \centering
 \vspace{-0.6cm}
  \caption{AMC results. The average accuracy on all SNRs of various models are listed in the table. Best result is highlighted in bold.}
  % \hspace{-20pt}
  \vspace{0.2cm}
\begin{tabular}{|c|c|c|}
\hline
AMC Model & Foundational Models &  Avg Accuracy   \\ \hline
MCFormer&  Transformer & 61.77\%  \\ \hline
LSTM  &  LSTM \cite{hochreiter1997long} & 62.51\%  \\ \hline
PET-CGDNN  & CNN + GRU \cite{dey2017gate} & 61.53\%  \\ \hline
MCLDNN  & CNN + LSTM & 61.86\%  \\ \hline
FEA-T & Transformer & 61.81\%  \\ \hline
LSRM & Transformer & 52.30\%  \\ \hline
HSRM & CNN &  66.05\%\\ \hline
MoE-AMC (our) & Transformer + CNN & \textbf{71.76\%}  \\ \hline
\end{tabular}

\label{result}
\end{table}

Fig. \ref{MoE-AMC} demonstrates the classification accuracies of MoE-AMC, as well as its two base models (LSRM and HSRM), at various SNR levels. The figure reveals that LSRM performs well under low SNRs. However, in certain low SNR scenarios, LSRM even surpasses its performance under high SNRs, indicating a certain level of overfitting to noise. On the other hand, HSRM yields suboptimal performance under low SNR conditions but performs well under high SNRs. MoE-AMC, combining the strengths of both models, achieves comparable performance to the previous leading models under high SNRs while exhibiting significant improvements over the previous leading models under low SNRs.

\section{Conclusion}
\label{sec:con}
In this study, we introduce MoE-AMC, a Mixture-of-Experts based model, for the task of Automatic Modulation Classification (AMC). Our proposed model takes into full consideration the variations in signals under different SNRs, and it combines the strengths of HSRM (a ResNet-based model) and LSRM (a Transformer-based model) to achieve superior performance in classifying modulation under both high and low SNR conditions. This model reduces the need for denoising preprocessing steps when classifying modulation signals in high-noise environments. The application of MoE techniques in AMC represents a novel contribution, showcasing its potential in handling diverse time series data. The scalability of the MoE framework also allows for future extensions to other time series-related tasks. 
%Overall, MoE-AMC presents a promising solution for accurate modulation classification in various noise environments.

\bibliographystyle{IEEEbib}
\bibliography{strings,refs}

\end{document}